# Track structure simulation of low energy electron damage to DNA using Geant4-DNA

**Edition: 2018/7/10**

Mojtaba Mokari[1, 2], Mohammad Hassan Alamatsaz[1], Hossein Moeini[1], Ali Akbar Babaei-Brojeny[1], and Reza Taleei[3]

*1- Department of Physics, Isfahan University of Technology, Isfahan 84156-83111, Iran*

*2- Department of Physics, Behbahan Khatam Alanbia University of Technology, Behbahan 63616-47189, Iran*

*3- Department of Radiation Oncology, University of Virginia, Charlottesville, VA, USA*

E-mail:  mokari@bkatu.ac.ir

**Abstract**

Due to the physical and chemical processes that are involved, interactions of ionizing radiations with cells lead to single- and double-strand breaks (SSB and DSB) and base damage to DNA cells. The damage may kill the cells or may be mis-repaired and lead to genetic diseases and cancers. Track structure Monte Carlo simulation of the DNA damage provides types of the damage and their frequencies. In the present work, to derive initial DNA damage, we used the Geant4-DNA code to simulate the physical, physico-chemical and chemical stages of interactions of incident beams of 100 eV– 4.5 keV electrons. By considering the direct damage of electrons and also the indirect hydroxyl radical damage to the DNA, in a simulation, simple and complex damages to SSB and DSB were investigated. Moreover, the yield of damage and the probability of types of DNA damage were evaluated. The results of these calculations were compared with the existing experimental data and the other simulations. For electrons with energies lower than 500 eV, there were differences between our results and published data which are basically due to the existing differences in the physical (electron ionization, excitation cross sections) and chemical models of Geant4-DNA, the chemical processes considered in the simulations, DNA geometry, and the selected parameters for damage threshold as compared to the other codes. In the present work, the effect of the threshold energy of the strand breaks was also evaluated.

**Introduction**

When ionizing radiation interacts with cells, the early and late biophysical effects are introduced. Initial effects include the effects from physical processes due to the ionization and excitation interactions as well as the effects of the chemical radicals. The damage to DNA, although not clinically recognizable, may give rise to genetic instability. Eventually, the short-term and also long-term effects of damage cause changes in the cellular structure and lead to cellular obstruction or cancer [1]. Understanding the mechanism of radiation damage involves knowledge of the spectrum of molecular damage that instigates initial biological lesions. Due to the differences in interactions and track patterns of various ionizing radiations, there are some differences in biological effects induced by such radiations. To infer the basic mechanisms of ionizing radiation interactions with cells, it is essential to determine the relevant physical, chemical, and biological parameters in cells. To study the effect of these parameters, relative data have been generated in structures of biological molecules such as DNA duplex and higher order structures. Especially, due to substantial evidence supporting the biological importance of clustered DNA damage, the DNA molecule is the likely candidate to consider. DNA damage includes single- and double-strand breaks (SSB and DSB) and is classified in the form of simple and complex breaks in cell nucleus. If the damage leads to a mis-repair or unrepair of DNA, especially DSB, this could give rise to the cell death [2-5].

Ionizing radiation damage to the DNA has been studied using both theoretical and experimental methods [6, 7]. A quantitative study of the parameters and effects of radiations has not yet been experimentally investigated by direct method [8]. Therefore, we studied the biophysical interactions by simulating the radiation transport in matter. The most successful track structure Monte Carlo codes for the physical (and chemical) simulations of radiation transport in matter are GEANT4-DNA [9], PITS [10], MCTS PARTRAC [11], and KURBUC [5] space-time code.

In calculating the damage and type of incident radiation, parameters such as energy, cross sections of interactions, $E_{ssb}$ threshold energy and the probability of indirect interactions of chemical radicals with DNA influence the results of SSB and DSB [12, 13]. There have been published results that only considered the direct damage induced by energy deposition in the DNA molecule [14-18]. Recently, there have been experimental-simulation studies performed with circular plasmid DNA by exploring

Auger-electron emitted from radionuclide [19]. In these studies, however, only direct damage by deposited energy in DNA using MCNP6 has been simulated. Some studies have been performed by Hahn *et al.* [17, 18] with experimental-simulation work with electron source and plasmid DNA using Geant4. In these studies, DNA damage was simulated only by direct effect of deposited energy. Pater *et al.* [16] also simulated electron beam in water medium using Geant4-DNA; in this work, however, DNA damage was measured only by the direct effect of deposited energy. Also, some previous works simulated DNA damage induced by both physical and chemical interactions [11, 20-22]. Meylan *et al.* [23] simulated fibroblast cell nucleus using Geant4-DNA with protons. Lampe *et al.* [24] effectively simulated the bacterial nucleus and studies the DNA damage from electrons and protons in a modelled full genome of an Escherichia coli cell using Geant4-DNA.

In this work, we used the Geant4-DNA (Geant4 version 10.3) code to simulate electrons with energies ranging from 100 eV to 4.5 keV in water and studied DNA damage. The aim was to calculate initial damage exerted on DNA by incident electrons using the Geant4-DNA code, which simulated both physical and chemical interactions and as such did a benchmarking of the Geant4-DNA performance for such calculations with previous existing experimental and simulation works. As well as, the $Yield_{SSB}$ and $Yield_{DSB}$, and complexity of the damage were reported. We also studied the effect of the threshold energy in the calculations.

**Materials and Methods**

*Monte Carlo Electron Simulation Considering Physical and Chemical Processes*

This work was performed using the Geant4-DNA (Geant4 version 10.3) code, which uses Monte Carlo technique for radiation-transport. The code follows the history of electron interactions in water by performing physical and chemical interactions. The Geant4-DNA code simulates physical interactions of primary and secondary electrons in the defined volume, and reports the interaction details such as energy transfer and coordinates of initial and secondary interactions [9, 25, 26]. The particles are tracked through the defined geometrical region and if a particle exits from the original mother volume, it is disregarded in the simulation.

The Geant4-DNA code is suitable for simulating the particle transport in water including physical and chemical interactions. In the current work and most other previous similar works, water cross sections were used. The cross sections used for physical interactions are the latest model used in Geant4 (version 10.3) and they have become more precise compared to previous models [27]. In the recent cross sections, all physical interactions such as elastic, ionization, excitation and Auger cascade processes are taken into consideration [9, 28]. The cross sections used in simulations of this work followed the original model of Geant4-DNA with 7.4 eV energy cutoff for electrons (electrons with lower energy than this value, deposit all their kinetic energy at this interaction point).

This study consisted of three stages. The first stage was the physical stage in which simulation of physical interactions of primary and secondary particles in water was considered until they reached the energy or geometrical cutoff. The second stage was the chemical stage which included the simulation of physico-chemical and chemical processes up to $10^{-9}$ seconds. The third stage was the damage formation stage in which a written algorithm determined types of damage in terms of complexity according to definition of damage spectra by Nikjoo *et al.* [12]. At the end of the physical stage, the coordinates and deposited energy during each step of the ionization and excitation interactions were derived from the code. Furthermore, at the end of the chemical stage, the coordinates of the produced radicals in the environment (water) were determined after $10^{-9}$ seconds. Table 1 displays the radicals and chemical interactions as well as the reaction rates, respectively, according to the Geant4 chemical model and experimental data. All the electron interactions including excitation, ionization, and cascade processes were simulated. The significance of studying chemical radicals and molecules has been proven in previous experimental studies [29, 30]. When the physical stage was terminated, the primary and secondary electrons were thermalized and they entered into the chemical stage ($10^{-15}$-$10^{-9}$ s). In this stage, the chemical radicals and molecules of $H_2O_2.H_2.e_{aq}.OH^-.OH^{\bullet}.H^+$ and $H^{\bullet}$ were produced in the environment. Then, chemical reactions occurred between molecules and radicals. In Table 1, these reactions are presented as they exist in the Geant4 code. To limit the time for this stage, the chemical stage duration was set to $10^{-9}$ seconds.

**Table 1.** Chemical interactions and radicals produced in the Geant4 [31] and experimental works (Exp.) [32].

| Reaction | Reaction Rate (Geant4) $(dm^3 mol^{-1} s^{-1})$ | Reaction Rate (Exp.) $(dm^3 mol^{-1} s^{-1})$ |
|---|---|---|
| $H_2 + OH^\bullet \rightarrow H^\bullet + H_2O$ | $4.17 \times 10^7$ | $4.5 \times 10^7$ |
| $OH^\bullet + OH^\bullet \rightarrow H_2O_2$ | $0.44 \times 10^{10}$ | $0.6 \times 10^{10}$ |
| $e_{aq}^- + e_{aq}^- + 2H_2O \rightarrow 2OH^- + H_2$ | $0.50 \times 10^{10}$ | $2.5 \times 10^{10}$ |
| $H^\bullet + H^\bullet \rightarrow H_2$ | $1.20 \times 10^{10}$ | $1.0 \times 10^{10}$ |
| $H_2O_2 + e_{aq}^- \rightarrow OH^- + OH^\bullet$ | $1.41 \times 10^{10}$ | $1.3 \times 10^{10}$ |
| $H^\bullet + OH^\bullet \rightarrow H_2O$ | $1.44 \times 10^{10}$ | $2.0 \times 10^{10}$ |
| $H_3O^+ + e_{aq}^- \rightarrow H^\bullet + H_2O$ | $2.11 \times 10^{10}$ | $1.7 \times 10^{10}$ |
| $H^\bullet + e_{aq}^- + H_2O \rightarrow OH^- + H_2$ | $2.65 \times 10^{10}$ | $2.5 \times 10^{10}$ |
| $OH^\bullet + e_{aq}^- \rightarrow OH^-$ | $2.95 \times 10^{10}$ | $2.5 \times 10^{10}$ |
| $H_3O^+ + OH^- \rightarrow 2H_2O$ | $14.3 \times 10^{10}$ | - |

*Simulation Geometry and Parameters*

Simulations were performed in a spherical water media with an isotropic electron source at the center of the sphere (100 nm radius). As mentioned, the primary and secondary electrons and chemical radicals were simulated using the Geant4-DNA code. The number of the primary electrons for each simulation was selected to reduce the uncertainty of the simulations below ±5%. For a proper distribution of DNA in the working volume sphere (WVS), and to reach a good statistical sampling, we had to sample a large number of DNAs (see Figure 1). The DNAs were produced through the µ-randomness method [33]. The sampling accuracy was tested using two criteria [34, 35]. In the first test, the ratio of energy deposition in the original sphere to its volume was compared to the ratio of energy deposition in the DNAs to their volumes. The criteria for a good sampling were the ratios of energy deposition within 5% uncertainty. In the second test, the mean specific energy frequency $\overline{Z}_f$ of the DNAs with the radius and length of 2.3 nm was calculated and compared to the deposited energy frequency $f(>0)$ [36, 37]. For the second test, the following criterion should be established: $f(>0) = \frac{1}{\overline{Z}_f}$. If the difference between the above tests were more than 5%, the sampling would be repeated with a larger number of DNAs [3].

*DNA Model Used in the Simulation*

Two types of DNA models have been employed earlier to model the DNA damage. Charlton *et al.* [38] and Nikjoo *et al.* [4, 12, 13, 39] used the B-DNA model. This model consists of a cylinder divided into sugar-phosphate and base regions without considering the details of atomic structures in oligonucleotides. The sugar-phosphate chains surround the center of a cylinder with a 10 Å -diameter and a 36-degree helical rotation. The DNA molecule diameter is 23 Å. Another common DNA model is Phosphodiester Groups (PDG) which consists of prisms with circular center bases used in the works of Bernal *et al.* [40, 41]. Friedland *et al.* [42, 43] also used the PDG model and defined the position of phosphor, oxygen, hydrogen and carbon atoms with van der Waals radius. Semenenko and Stewart [44, 45] instead of using the DNA model, used the genome distances in the MCDS code.

In this work, the DNA model used was a 216 bp long double helix B-DNA (equivalent to 73.44 nm and consisting of 432 nucleotides). The B-DNA model is one of the most common kinds of double helix DNA types found in cells [46-48]. The length of the DNA model in this work was 216 bp, and its diameter was 23 Å and consisted of 432 nucleotides. Each nucleotide consisted of a sugar-phosphate backbone and a base group of four species of Adenine, Cytosine, Guanine, and Thymine.

*Direct Interactions and Threshold Energy $E_{ssb}$*

DNA damage induced by ionizing radiations is direct or indirect. For a direct damage, a threshold energy ($E_{ssb}$) is determined. $E_{ssb}$ is the least amount of energy required to cause break in each strand of DNA. The possibility of direct damage might be determined through the comparison of $E_{ssb}$ in a nucleotide with quantities such as the total deposited [13, 49], maximum deposited [50], total transferred and maximum transferred energy [40, 41]. In the present work, we studied the total deposited energy (in all events) for examining the possibility of direct damage. For $E_{ssb}$, different values have been chosen in different works. The most used threshold energy is 17.5 eV [13, 23, 38] and 10.79 eV [15, 16, 49]. For DNA damage simulations, where indirect damage by chemical radicals was not considered, the threshold energy was chosen as $E_{ssb}$ = 10.79 eV. In this work, given the chemical radicals effects and indirect damage yield, the threshold energy was chosen as 17.5 eV. However, 17.5 eV has been found to be an appropriate threshold energy given by the experimental findings of the spectrometry of Auger electrons and I-125 experiments [51-54]. If the total energy deposition in the nucleotide sugar-phosphate groups is equal or more than the $E_{ssb}$, strand break (SB) occurs.

*Indirect Interaction and Hydroxyl Radical Damage*

In the chemical stage, chemical radicals and molecules interact according to Table 1. The $e_{aq}$. $OH^\bullet$. and $H^\bullet$ radicals interact with the DNA sugar and base groups of nucleotides. The likelihood of hydroxyl radical interaction is much more compared to the other two radicals: $e_{aq}$ and $H^\bullet$ [55]. Thus, the hydroxyl radical share in causing damage in DNA is investigated. Hydroxyl radical interacts with sugar-phosphate groups or nucleobases and produces sugar or base radicals [56]. The probability of hydroxyl radical interacting with the base and sugar-phosphate is 80% and 20%, respectively. Therefore, the sugar radicals produced due to the interaction of hydroxyl with sugar-phosphate lead to SB with a 65% probability. Consequently, the probability of SB damage (indirect damage) due to the interactions of hydroxyl radical with DNA nucleotides is equal to 13% ($P_{OH}$= 0.13) [57].

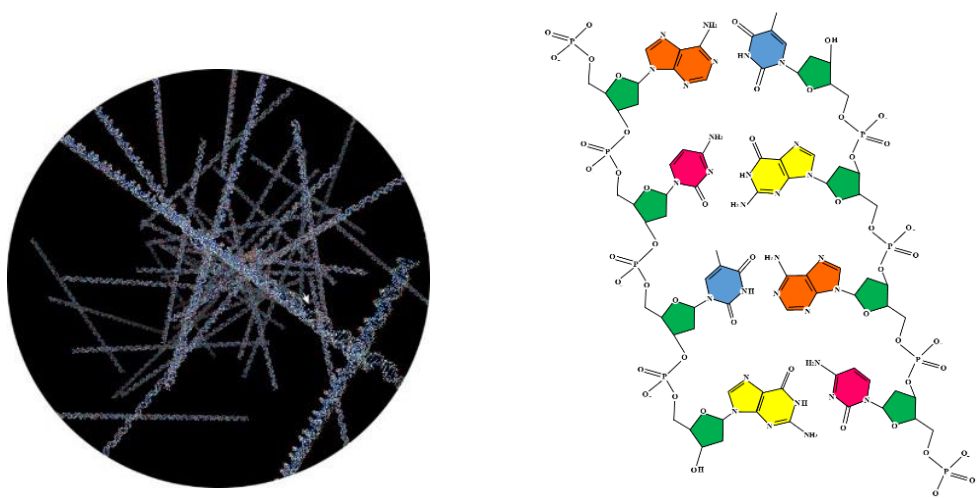

**Figure 1.** 50 DNA segments randomly distributed within the spherical water environment (left) drawn using VMD software (http://www.ks.uiuc.edu/Research/vmd/). On the right, a schematic view of part of the DNA molecule containing base, sugar, and phosphate chains is shown.

*Damage Mechanism and its Categorization*

We developed a C++ program to sample a large number of B-DNAs in the WVS. We also developed a Python program to compute the damage distances to find the closest nucleotide to the energy deposition points and the coordinates of hydroxyl radicals. The derived positions of the hydroxyl radicals were checked in our algorithm to see whether they would fall within the volume of any imaginary cylinder of (8 + 2.3) nm diameter, with its longitudinal axis coinciding with the axis of the DNA cylinder of 2.3 nm diameter. Having the $E_{ssb}$ and $P_{OH}$, then we specified the types of the DNA damage. To perform the sampling method mentioned in the previous section (Simulation Geometry and Parameters), we chose a large number of DNAs. These samples were distributed randomly in the WVS in different directions. The direct or indirect damage induced to the opposite strands of the DNA within less than 10 bp is considered as DSB. The different types of the DNA damage are divided into two categories of simple and complex.

Complex damage includes SSB+, DSB+, and DSB++. Figure 2 shows different types of DNA damage. To categorize damage, various models have been presented by different authors such as Friedland *et al.* [11, 49], Bernal *et al.* [40, 41, 58], Nikjoo *et al.* [5, 12], Charlton *et al.* [34] and Pater *et al.* [16]. In this work, the damage was categorized using Nikjoo's definition. In Nikjoo's definition the damage is named accordingly as DSB++, DSB+, DSB, SSB+, 2SSB, SSB and NB (no break).

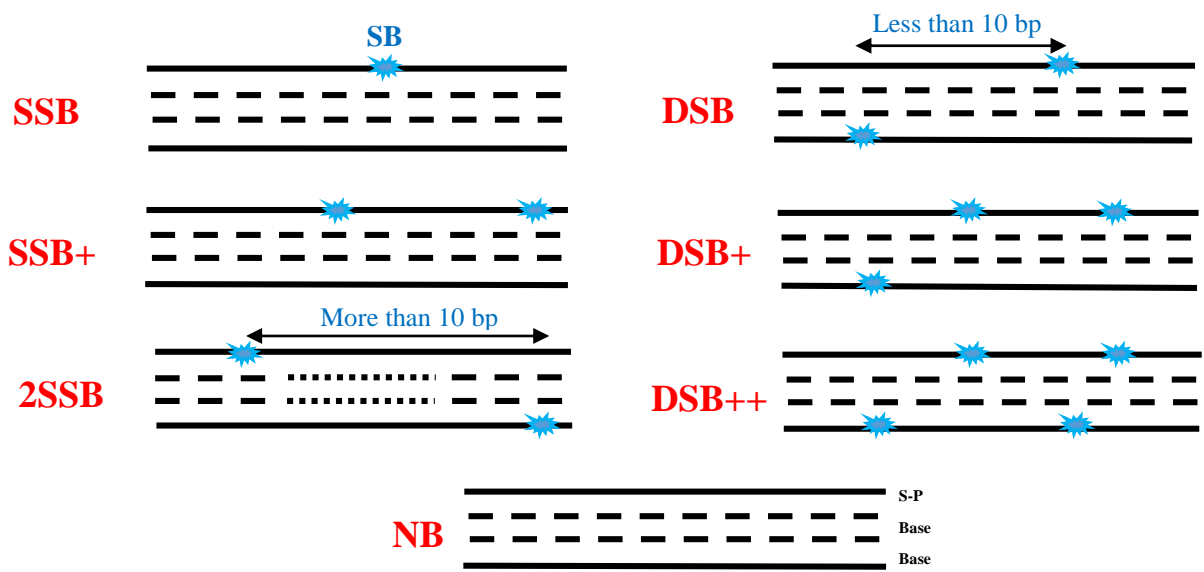

**Figure 2.** Model of the DNA damage induced by direct energy deposition and reaction of hydroxyl radicals. For ease of observation, the DNA is shown as four untwisted linear lines. The solid lines at the top and bottom represent the sugar-phosphate (S-P) backbone; the two dash lines represent the bases. A '*' represents an SB in DNA. If two '*'s are on opposite strands within 10bp of each other, it will be considered a *DSB*. If two SSBs are more than 10 bp apart, it is denoted by *2SSB*, and if two SSBs are within 10 bp apart, but on the same strand of DNA, it is denoted by *SSB+*. A double strand break accompanied by one (or more) additional single strand break within 10 bp separation is denoted by *DSB+*. More than one double strand break on the segment either within the 10 bp separation or further apart is denoted by *DSB++*. The *NB* (no break) category refers to a DNA without any SBs [5, 59].

**Results**

For simulating direct and indirect damage to DNA, we assumed a water sphere. The electron source was located in the center and emitted electrons in random directions. The electron energies for the sources were 100 eV, 300 eV, 500 eV, 1 keV, 1.5 keV, and 4.5 keV. The physical and chemical interactions of the electrons in the water environment were simulated with $10^3 - 10^4$ history. The direct and indirect DNA damage induced by electrons was calculated using an algorithm written in the Python program, given a threshold energy of $E_{ssb}$= 17.5 eV (or 30 eV) and hydroxyl radical interaction probability of $P_{OH}$ = 0.13. The damage was categorized and studied according to Nikjoo's method presented in Figure 2. In Table 2, the calculated relative yields of different types of strand breaks have been displayed for the threshold energy of $E_{ssb}$ = 17.5 eV and hydroxyl radical interaction probability of $P_{OH}$ = 0.13. When damage occurs on sugar-phosphate, it can lead to simple damage (SSB and DSB) or complex damage (DSB++, DSB+, SSB+). There are other types of complex damage categorized as $SSB_c$ (= $SSB^+$ + 2SSB) and $DSB_c$ (= $DSB^+$ + $DSB^{++}$) [4], which were calculated in this work and presented in Tables 2 and 4. The results showed that the probability of $SSB_c$ of energies ranging from 100 eV to 1 keV increased and then decreased. Moreover, the probability of $DSB_c$ for energies from 300 eV to 4.5 keV decreased. The minimum and maximum $Yield_{DSB}$ occurred at 4.5 keV and 500 eV energies, respectively. Moreover, the least and most $Yield_{SSB}$ values were at 1.5 keV and 500 eV energies, respectively. In Figure 3, the relative damage yields predicted by this work is compared with the results of Nikjoo *et al.* [4, 13] using the CPA100 code and also with those of Taleei *et al.* [21] using the KURBUC$_{liq.}$ code. The probability of simple SSB calculated in this work for energies ranging from 100 eV to 500 eV (Figures 3-a, b, and c) was less than Nikjoo and Taleei's calculations, and for energies ranging from 1 keV to 4.5 keV (Figures 3-d, e, and f) was more than Nikjoo and Taleei's results. Moreover, the probability of the DSB damage, especially complex DSB, was more than Nikjoo and Taleei's studies. However, the trend of the probability of simple and complex damage yields as a function of energy is similar to the Nikjoo and Taleei's results.

**Table 2.** Relative yield of the strand breaks classified by damage complexity with $E_{ssb} = 17.5$ eV and $P_{OH} = 0.13$

| Energy eV | No Break % | SSB % | SSB+ % | 2SSB % | DSB % | DSB+ % | DSB++ % | $SSB_c$ % | $DSB_c$ % | $Y_{SSB}$ Gy$^{-1}$Gbp$^{-1}$ | $Y_{DSB}$ Gy$^{-1}$Gbp$^{-1}$ |
|---|---|---|---|---|---|---|---|---|---|---|---|
| 100 | 66.72 | 21.94 | 3.55 | 2.63 | 3.68 | 1.36 | 0.11 | 21.98 | 28.55 | 81.62 | 10.25 |
| 300 | 45.41 | 19.76 | 5.16 | 5.67 | 7.65 | 9.89 | 6.77 | 35.41 | 68.14 | 101.46 | 28.91 |
| 500 | 38.81 | 22.26 | 4.31 | 9.76 | 9.55 | 10.39 | 4.89 | 38.77 | 61.54 | 114.01 | 29.55 |
| 1000 | 37.04 | 29.01 | 3.83 | 15.59 | 9.58 | 4.11 | 0.83 | 40.10 | 34.01 | 104.08 | 16.24 |
| 1500 | 42.78 | 34.03 | 3.24 | 12.83 | 5.1 | 1.85 | 0.17 | 32.07 | 28.47 | 77.16 | 7.12 |
| 4500 | 66.35 | 26.81 | 1.13 | 4.03 | 1.44 | 0.22 | 0.03 | 16.16 | 14.39 | 109.61 | 4.68 |

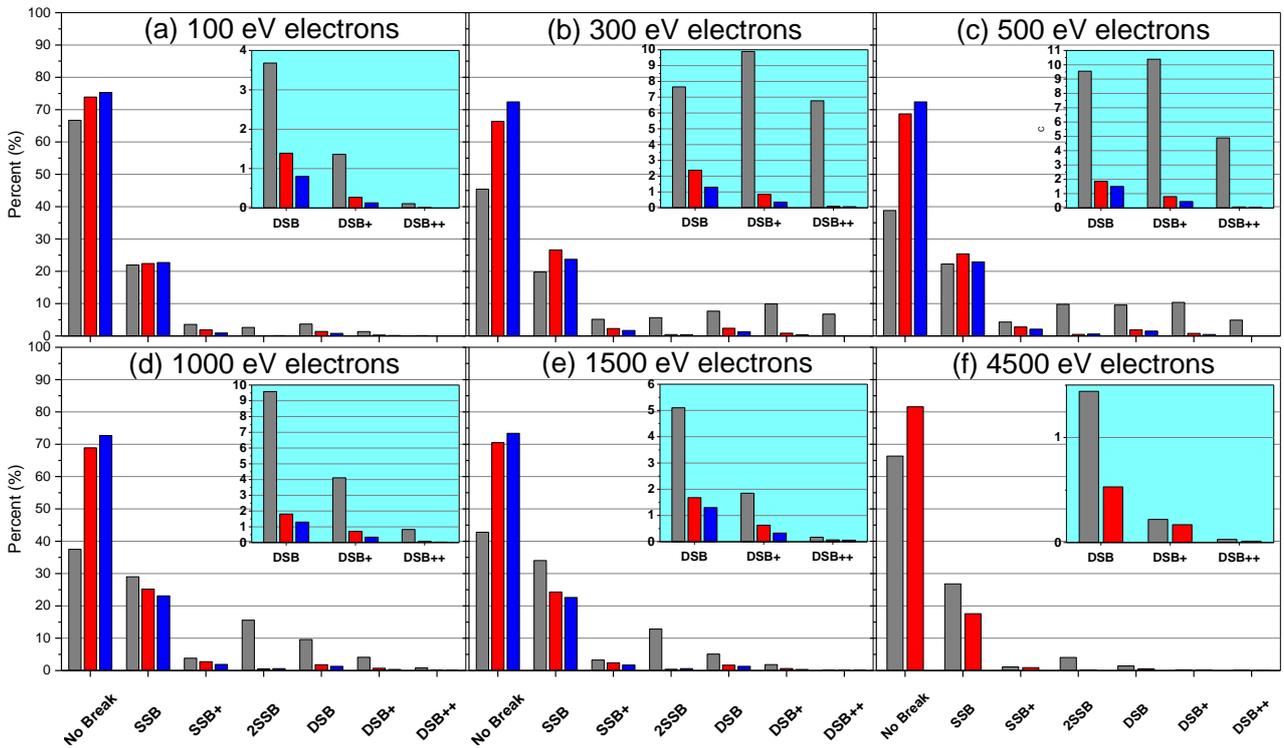

**Figure 3.** Comparison of the DNA damage spectra predicted by this work (grey), Nikjoo's (red), and Taleei's (blue) for 100 eV (a), 300 eV (b), 500 eV (c), 1000 eV (d), 1500 eV (e), and 4500 eV (f) electrons with $E_{ssb} = 17.5$ eV and $P_{OH} = 0.13$

In Figure 4, the Yield$_{SSB}$ and Yield$_{DSB}$ values of the current work and previous experimental and simulation works are compared. In this figure, the yield values for our simulation DSB damage are compared to Yield$_{DSB}$ in de Lara *et al.* [7], which was measured with Chinese hamster cells. Moreover, our results are compared to simulations of Nikjoo *et al.* [4, 12, 13] using the CPA100 code, Semenenko and Stewart [44, 45] using the MCDS code, Bernal and Liendo [40] using the PENELOPE code, and Friedland *et al.* [42, 49] using the PARTRAC code. In Figure 4-a, the Yield$_{DSB}$ values were compared with experimental results of de Lara and previous simulation works. The relative difference of $Yield_{DSB}$ between our results and de Lara' results was 11.15% and 55.68% at 1 keV and 4.5 keV, respectively. The $Yield_{DSB}$ in this study at 1 keV and higher energies are closer to those of the other simulation works. At energies of about 500 eV and 300 eV in Figure 4-a, there were differences between various studies. The relative difference of $Yield_{DSB}$ between our simulation and Nikjoo's results was between 3.54% at 100 eV and 123.86% at 500 eV. Moreover, the relative difference of our results and Semenenko's was between 26.31% at 1 keV and 59% at 100 eV. The $Yield_{DSB}$ relative difference at 1.5 keV was 48.33% in Benal's simulation and 16.24% in Friedland's simulation. Figure 4-b shows that the trend of changes was similar to Nikjoo's results. The computed $Yield_{SSB}$ values in the current study were close to those obtained in the works of Friedland and Bernal.

In order to study the effect of the threshold energy of $E_{ssb}$, we calculated the simple and complex SSB and DSB values at threshold energies of 12.6, 15.0, 17.5, 21.1, 30.0 which were the most commonly used threshold energies in previous works. For this purpose, at 300 eV energy, assuming indirect interaction was not present, we calculated the ratio of the total number of DSB to the total SSB ($SSB_{all} = SSB + SSB^+ + 2 \times (2SSB + DSB + DSB^+ + DSB^{++})$ and $DSB_{all} = DSB + DSB^+ + DSB^{++}$ [38]). This test was performed on $10^4$ molecules of DNA in the WVS. The $SSB_{total}/DSB_{total}$ ratio fluctuates from 3.68 to 9.03. Table 3 lists the damage calculation yields at different threshold energies. As seen in Table 3, by increasing the threshold energy $E_{ssb}$, the ratio of $SSB_{total}/DSB_{total}$ increases. It can be seen that the induced DNA damage is strongly dependent on $E_{ssb}$. In the Nikjoo *et al.* [12] the ratio of $SSB_{total}/DSB_{total}$ is approximated to 8.5, for the threshold energy of 17.5 eV, and in the higher threshold energy, the growth of this rate is found suddenly to be significant. Consequently, it seems that in the current Geant4 model for electrons, with the choice of larger threshold energy, yields values are closer to results of other experimental and simulation works. For this purpose, we examined the threshold energy of 30.0 eV (ratio= 9.03 in Table 3 that is close to the amount of 8.5 in Nikjoo *et al.* [12]), and it is one of the most commonly used threshold energies in the previous works.

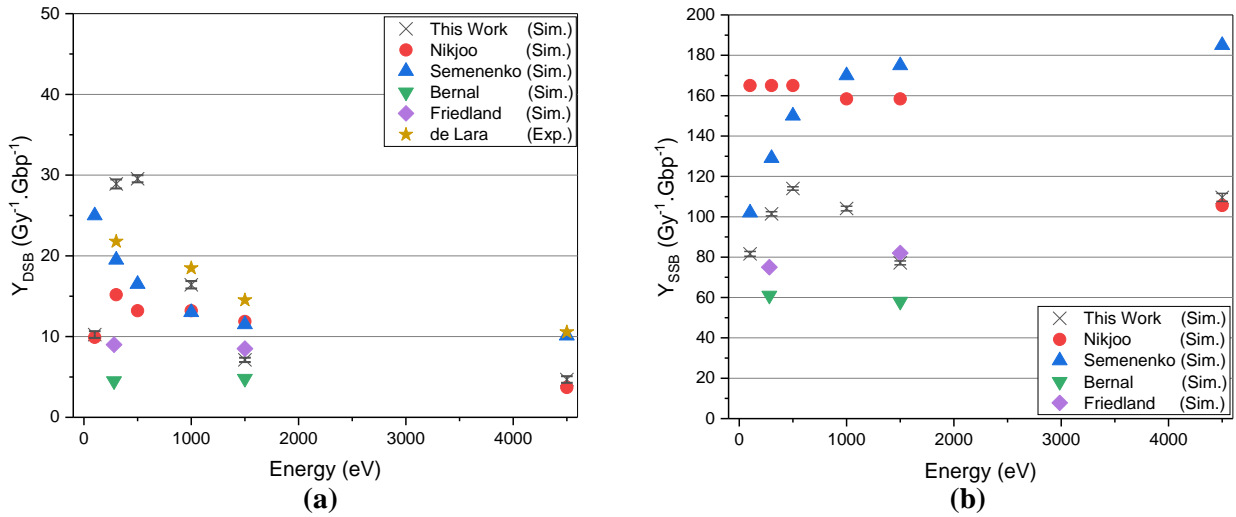

**Figure 4.** Comparison of DSB (a) and SSB (b) yield values with those of the experimental and simulations results.

**Table 3.** Threshold energy dependence for direct damage with zero hydroxyl radical activation probability

| Threshold Energy | SSB | SSB+ | 2SSB | DSB | DSB+ | DSB++ | Total SSB/Total DSB |
|---|---|---|---|---|---|---|---|
| **12.6 eV** | 1327 | 491 | 258 | 473 | 541 | 376 | 3.68 |
| **15.0 eV** | 1069 | 324 | 134 | 257 | 295 | 144 | 4.39 |
| **17.5 eV** | 943 | 262 | 133 | 246 | 204 | 75 | 4.80 |
| **21.1 eV** | 938 | 261 | 133 | 236 | 199 | 70 | 4.90 |
| **30.0 eV** | 616 | 99 | 50 | 65 | 49 | 2 | 9.03 |

Table 4 shows the calculated relative yields of different types of strand breaks, considering $E_{ssb}$ = 30.0 eV and $P_{OH}$ = 0.13. In addition, Figure 5 presents the relative damage yields predicted by current study for the threshold energies of 30.0 eV and 17.5 eV with an equal indirect damage probability ($P_{OH}$ = 0.13). In this figure it is observed that with the increase of $E_{ssb}$, the probability of complex DSB damage decreases. Also, the probability of hits without the NB damage increases. As the threshold energy increases, due to reduction in multi strand breaks on a DNA, in all figures, the probability of simple and complex DSBs decreases. Moreover, the probability of SSB increases at energies equal to or less than 1 keV (Figures 5-a, b, c, and d). However, with increasing energy (Figures 5-e, and f) due to a reduction in the overall share of SBs through the threshold energy, SSB probability decreases. It is apparent that the results may have been dependent on parameter assumptions in the simulation. In Figure 6-a and c, the yield values for the threshold energies of 30.0 eV and 17.5 eV with equal indirect damage probability ($P_{OH}$ = 0.13) are compared. Comparing the results corresponding to threshold energy of 30.0 eV and 17.5 eV, Yield$_{SSB}$ and Yield$_{DSB}$ decrease. Yield$_{DSB}$ for either of the threshold energies decreases as a function of primary electron energy. Moreover, the highest drop rate was observed for 4.5 keV and as the energy increased, the relative reduction of the yield also increased. In Figure 6-b and d, it is seen that with the increase in the threshold energy to 30.0 eV, Yield$_{SSB}$ approaches the results of Bernal and Friedland. Also, Yield$_{DSB}$ values in energies below 500 eV are closer to experimental results. The trend of the yield results is similar to those of Nikjoo's results, especially in yield results for energies lower than 500 eV.

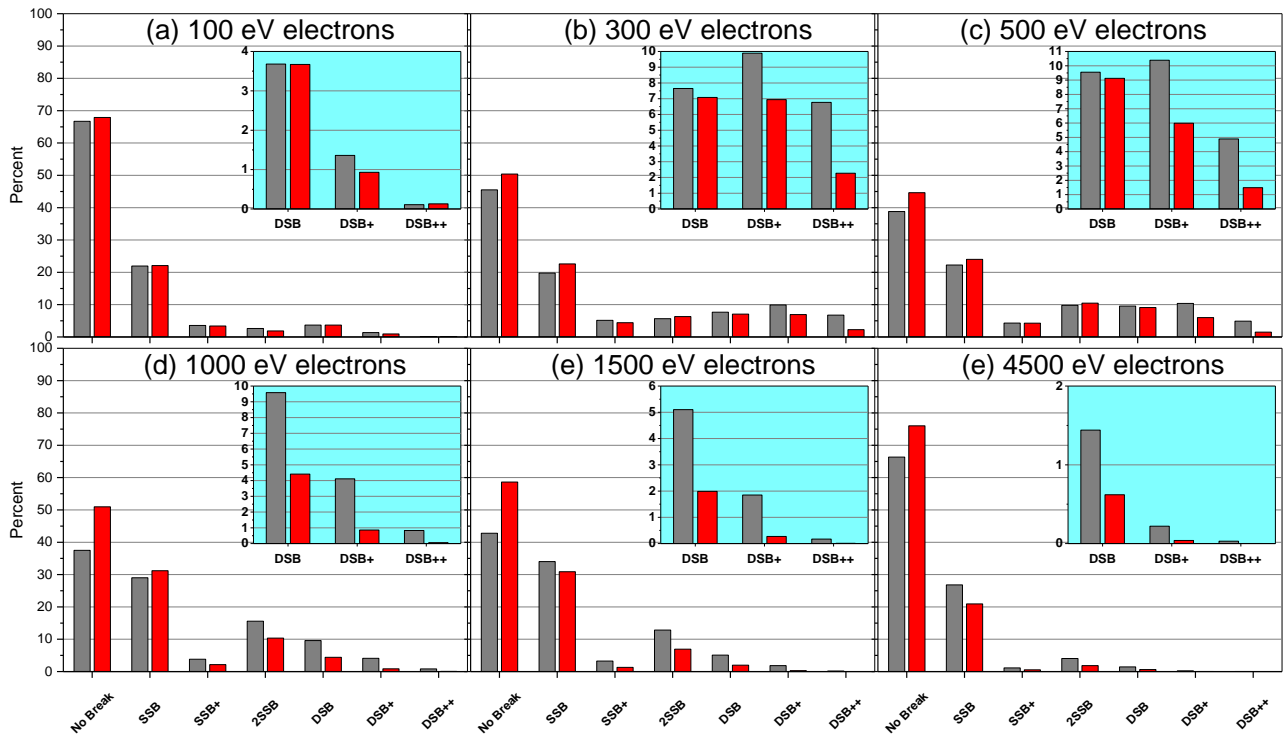

**Figure 5.** Comparison of the DNA damage spectra predicted by this work with $E_{ssb}$ = 17.5 eV (grey) and $E_{ssb}$ = 30.0 eV (red) with $P_{OH}$ = 0.13 for 100 eV (a), 300 eV (b), 500 eV (c), 1000 eV (d), 1500 eV (e), and 4500 eV (f) electron

**Table 4.** Relative yield of the strand breaks classified by damage complexity with $E_{ssb}$ = 30.0 eV and $P_{OH}$ = 0.13

| Energy eV | No Break % | SSB % | SSB+ % | 2SSB % | DSB % | DSB+ % | DSB++ % | SSB$_c$ % | DSB$_c$ % | Y$_{SSB}$ Gy$^{-1}$Gbp$^{-1}$ | Y$_{DSB}$ Gy$^{-1}$Gbp$^{-1}$ |
|---|---|---|---|---|---|---|---|---|---|---|---|
| **100** | 67.91 | 22.10 | 3.40 | 1.87 | 3.66 | 0.93 | 0.13 | 19.22 | 22.54 | 79.52 | 9.72 |
| **300** | 50.40 | 22.60 | 4.40 | 6.31 | 7.08 | 6.94 | 2.27 | 32.15 | 56.53 | 89.81 | 20.26 |
| **500** | 44.64 | 24.01 | 4.28 | 10.46 | 9.12 | 5.99 | 1.49 | 38.04 | 45.06 | 99.52 | 20.04 |
| **1000** | 50.97 | 31.20 | 2.16 | 10.34 | 4.41 | 0.86 | 0.06 | 28.61 | 17.24 | 71.88 | 5.92 |
| **1500** | 58.63 | 30.87 | 1.31 | 6.92 | 1.99 | 0.27 | 0.01 | 21.04 | 12.33 | 50.22 | 2.25 |
| **4500** | 76.03 | 20.94 | 0.55 | 1.83 | 0.62 | 0.04 | 0.001 | 10.20 | 6.00 | 71.31 | 1.77 |

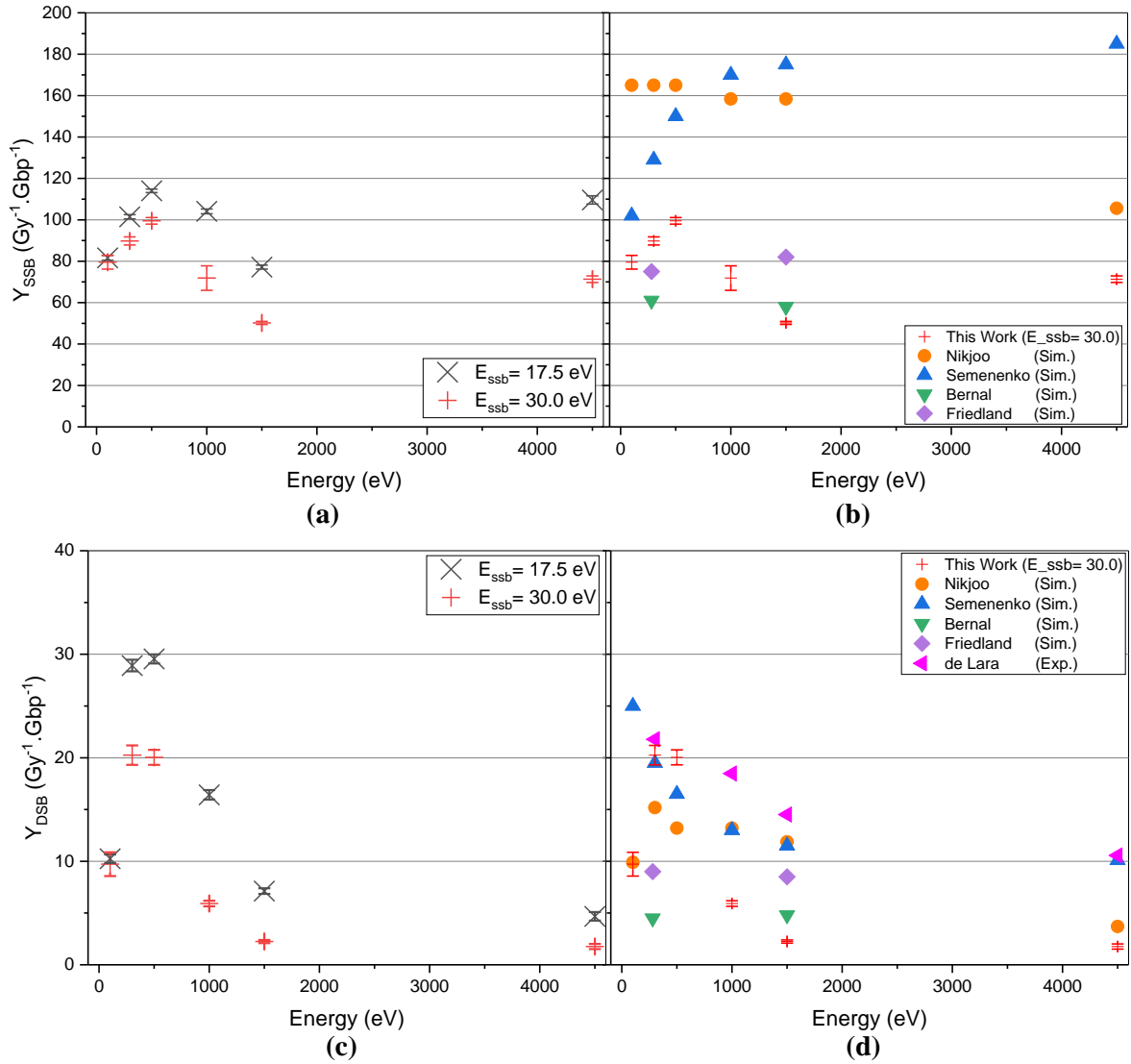

**Figure 6.** Comparison of SSB (a) and DSB (c) yield values for $E_{ssb}$ = 17.5 eV and $E_{ssb}$ = 30.0 eV with $P_{OH}$ = 0.13. Also comparison of SSB (b) and DSB (d) yield values with those of the experimental and simulations results for $E_{ssb}$ = 30.0 eV with $P_{OH}$ = 0.13

**Discussion**

In this work, a large number of electron events were transported from the center of the water sphere. The primary electron interactions were simulated by the Geant4-DNA code. Subsequently, the yield of damage in the DNA samples was calculated, in a process we referred here as damage formation stage. For the physical stage, the threshold energy for recording a hit as a break was considered to be 17.5 and 30.0 eV. Same value has been used in the simulations by Nikjoo *et al.* [12] and Taleei *et al.* [21] where they simulated B-DNAs using the CPA100 and KURBUC codes. Using the PARTRAC code, Friedland

*et al.* [42, 49] have investigated a threshold variation between 5 and 37.5 eV, implementing a linear acceptation probability (a linear increasing of the probability from zero, for a deposited energy less than 5 eV, to 1 when it exceeds 37.5 eV) for direct damage [43]. Friedland *et al.* have implemented a basic chromatin fiber element including 30 nucleosomes and an ideal arrangement of chromatin fiber rods in rhombic loops forming a rosette-like structure of 0.5 Mbp genomic length. We have adopted the interaction probability of the hydroxyl radicals 0.13 which is the same as in Nikjoo and Friedland's works. Like in Nikjoo *et al.* [12] and Taleei *et al.* [21], we limited the chemical stage simulation time to 1 ns for the interaction of hydroxyl radicals with DNA. In our simulations, we did not specifically model the scavenging reactions that decrease the number of the existing hydroxyl radicals for damaging the DNA, whereas Friedland *et al.* [42, 60] has taken into account the scavenging of the chemical species at each time step due to random absorption of the radicals and as such considered an appreciably longer chemical stage simulation time of 10 ns.

The differences in the yield values observed in Figure 4 and 6 are primarily due to differences in the physical (ionization, excitation cross sections) and chemical models of Geant4-DNA, the chemical processes considered in the simulations, and DNA geometry [61, 62]. For example, there are differences between the excitation cross sections of the CPA100 and Geant4-DNA codes which are shown to be about an order of magnitude different for electron energies higher than 100 eV [62]. The cross sections of the CPA100 ionization model are in closer agreement to experimental data as compared to the other models [63]. Although for electrons with energies higher than 100 eV, which ionization is known to be the most important process, the ionization cross sections in Geant4-DNA are in a reasonable agreement with the ones in CPA100 [62]. It is also worth to mention that the maximum of the total excitation cross sections in Geant4-DNA is shown to be lower than the one from the PARTRAC code [64].

Moreover, reaction rates listed in Table 1 for the Geant4 chemical model and experimental data, it can be observed that the chemical reaction rates of the hydroxyl radicals with other molecules and radicals (including other hydroxyl radicals) are less in Geant4-DNA. Moreover, the production rate values of the hydroxyl radicals are larger in Geant4-DNA as compared to the other experimental values (see the fifth row of Table 1). Therefore, in the Geant4 code, more hydroxyl radicals reacted in the environment and the share of indirect damage was higher. At 500 eV and close to 300 eV energies

(Figures 4-a and b), due to the models of electron interactions and chemical reactions in the Geant4-DNA code, the deposited energy of ionization and excitation was closer to the produced hydroxyl radicals after electron full-stop and thus, caused more DSBs, especially complex DSB (Figure 3 and Table 2). This led to an increase in Yield$_{DSB}$ and decrease in Yield$_{SSB}$.

Also in our simulation, the action of hydroxyl radical interacting with base and base damage was not taken into consideration. The latter effect was also ignored in other published simulations; however, they can affect the SB damage yield [65]. Additionally, the uncertainty of the simulations increases at lower electron energies [5].

According to the results of Figure 4 at energies above 500 eV, especially in DSB yields, our results were close to the experimental and simulation works, taking into account the threshold energy of 17.5 eV. Using the threshold energy of 30.0 eV (Figure 6-b and d), for primary electrons with energies lower than 500 eV, the yield results were closer to the experimental and simulation results. Therefore, with the default Geant4-DNA model with primary electrons lower than 500 eV and threshold energy higher than the usual 17.5 eV ($E_{ssb}$= 30.0 eV), our simulation approximates the predicted results. In the next works, we will use the CPA100 cross sections in Geant4-DNA, and because of their proximity to experimental values, we are trying to obtain more accurate results.

**Conclusions**

The main purpose of this work was to simulate the frequency of simple and complex damages in a B-DNA model using the Geant4-DNA code and as such did a benchmarking of the Geant4-DNA performance with some other works. Using the track structure simulation tools, we were able to simulate energy deposition of the physical processes and chemical reactions of hydroxyl radicals in the DNA model. This work was performed by simulating physical and chemical stages using Geant4-DNA and an analysis algorithm using Python program. In this work, we used large number of electron events that were randomly transported from the water sphere center with energies ranging from 100 eV to 4.5 keV. Then, the probability of simple and complex damages as well as that of the Yield$_{SSB}$ and Yield$_{DSB}$ was calculated. Further, the effect of $E_{ssb}$ amounts in the calculations was studied. These calculations showed the dependence of the direct DNA damage with the threshold energy. Taking into account the threshold energy of 30.0 eV, the yield results were closer to the experimental values for primary electrons with energies lower than 500 eV. Further, we compared the results of this work with the corresponding simulations and experimental DNA damage results induced by electrons. There were differences between the results of this work and those of other works, especially at energies below 500 eV. We believe that the reasons for the differences are due to the difference in the physical and chemical models of Geant4-DNA with other codes, the type of chemical processes considered in simulation, DNA geometry, and the selected parameters for damage threshold.

**Acknowledgments**

The authors gratefully acknowledge the Sheikh Bahaei National High Performance Computing Center (SBNHPCC) for providing computing facilities and time. The SBNHPCC is supported by Isfahan University of Technology (IUT).

**References:**

[1] Nikjoo H and Uehara S 2004 *Track Structure Studies of Biological Systems, in Charged Particle and Photon Interactions with Matter: Chemical, Physiochemical, and Biological Consequences with Applications*, Mozuumder A and Hatano Y Eds. (New York, Marcel Dekker)

[2] Goodhead D T and Nikjoo H 1989 Track structure analysis of ultrasoft X-rays compared to high- and low-LET radiations *Int. J. Radiat. Biol.* **55** 513-529

[3] Nikjoo H, Goodhead D E, Charlton D E and Paretzke H G 1991 Energy deposition in small cylindrical targets by monoenergetic electrons *Int. J. Radiat. Biol.* **60** 739-756

[4] Nikjoo H, O'Neill P, Wilson W E and Goodhead D T 2001 Computational Approach for Determining the Spectrum of DNA Damage Induced by Ionizing Radiation *Radiat. Res.* **156** 577–583

[5] Nikjoo H, Emfietzoglou D, Liamsuwan T, Taleei R, Liljequist D and Uehara S 2016 Radiation track, DNA damage and response-a review *Rep. Prog. Phys.* **79** 116601

[6] Frankenberg D, Frankenberg-Schwager M, Bloecher M and Harbich R 1981 DNA Double Strand Breaks as the Critical Lesion in Yeast Cells Irradiated with Sparsely or Densely Ionising Radiation under Oxic or Anoxic Conditions *Radiat. Res.* **88** 524–532

[7] de Lara C M, A Hill M, Papworth D and O'Neill P 2001 Dependence of the Yield of DNA Double-strand Breaks in Chinese Hamster V79–4 Cells on the Photon Energy of Ultrasoft X Rays *Radiat. Res.* **155** 440–448

[8] Nikjoo H, Uehara S and Emfietzoglou D 2012 *Interaction of Radiation with Matter* (Boca Raton, FL: CRC Press)

[9] Incerti S, Ivanchenko A, Karamitros M, Mantero A, Moretto P, Tran H N and et al. 2010 Comparison of GEANT4 very low energy cross section models with experimental data in water *Medical Physics* **37** 4692–4708

[10] Wilson W E and Nikjoo H 1999 A Monte Carlo code for positive ion track simulation *Radiat. Environ. Biophys.* **38** 97–104

[11] Friedland W, Dingfelder M, Kundrát P and Jacob P 2011 Track structures, DNA targets and radiation effects in the biophysical Monte Carlo simulation code PARTRAC *Mutation Research/Fundamental and Molecular Mechanisms of Mutagenesis* **711** 28–40

[12] Nikjoo H, O'Neill P, Goodhead D and Terrisol M 1997 Computational Modelling of Low-energy Electron-induced DNA *Int. J. Radiat. Biol.* **71** 467-483

[13] Nikjoo H, Bolton C E, Watanabe R, Terrissol M, O'Neill P and Goodhead D T 2002 Modelling of DNA damage induced by energetic electrons (100 eV to 100 keV) *Radiation Protection Dosimetry* **99** 77-80

[14] Martin F, Burrow P D, Cai Z, Cloutier P, Hunting D and Sanche L 2004 DNA Strand Breaks Induced by 0–4 eV Electrons: The Role of Shape Resonances *Physical Review Letters* **93** 068101

[15] Incerti S, Champion C, Tran H N, Karamitros M, Bernal M, Z Francis, Ivanchenko V, Mantero A and Members of the Geant4-DNA collaboration 2013 Energy deposition in small-scale targets of liquid water using the very low energy electromagnetic physics processes of the Geant4 toolkit *Instrum. Methods Phys. Res., Sect. B* **306** 158–164

[16] Pater P, Seuntjens J, El Naqa I and Bernal M 2014 On the consistency of Monte Carlo track structure DNA damage simulations *Medical Physics* **41** 121708

[17] Hahn M B, Meyer S, Schroter M A, Seitz H, Kunte H J, Solomun T and Sturm H 2017 Direct electron irradiation of DNA in a fully aqueous environment. Damage determination in combination with Monte Carlo simulations *Phys. Chem. Chem. Phys.* **19** 1798-1805

[18] Hahn M B, Meyer S, Kunte H J, Solomun T and Sturm H 2017 Measurements and simulations of microscopic damage to DNA in water by 30 keV electrons: A general approach applicable to other radiation sources and biological targets *Physical Review E* **95** 052419

[19] Di Maria S, Belchior A, Pereira E, Quental L, Oliveira M C, Mendes F and Lavrado J 2017 Dosimetry assessment of DNA damage by Auger-emitting radionuclides: Experimental and Monte Carlo studies *Radiation Physics and Chemistry* **140** 278–282

[20] Taleei R and Nikjoo H 2012 Repair of the double-strand breaks induced by low energy electrons: A modelling approach *International Journal of Radiation Biology* **88** 948-953

[21] Taleei R, Girard P M, Sankaranarayanan K and Nikjoo H 2013 The Non-homologous End-Joining (NHEJ) Mathematical Model for the Repair of Double-Strand Breaks: II. Application to Damage Induced by Ultrasoft X Rays and Low-Energy Electrons *Radiation Research* **179** 540-548

[22] Friedland W, Schmitt E, Kundrát P, Dingfelder M, Baiocco G, Barbieri S and Ottolenghi A 2017 Comprehensive track-structure based evaluation of DNA damage by light ions from radiotherapy relevant energies down to stopping *Sci. Rep.* **7** 45161

[23] Meylan S, Incerti S, Karamitros M, Tang N, Bueno M, Clairand I and Villagrasa C 2017 Simulation of early DNA damage after the irradiation of a fibroblast cell nucleus using Geant4-DNA *Scientific Reports* **7** 11923

[24] Lampe N, Karamitros M, Breton V and et al. 2018 Mechanistic DNA damage simulations in Geant4-DNA Part 2: Electron and proton damage in a bacterial cell *Physica Medica* **48** 146–155

[25] Incerti S et al. 2010 The Geant4-DNA project, *International Journal of Modeling, Simulation, and Scientific Computing* **1** 157–178

[26] Bernal M A and et al. 2015 Track structure modeling in liquid water: A review of the Geant4-DNA very low energy extension of the Geant4 *Physica Medica* **31** 861–874

[27] Kyriakou I, Šefl M, Nourry V and Incerti S 2016 The impact of new Geant4-DNA cross section models on electron track structure *Journal of Applied Physics* **119** 194902

[28] Champion C, Incerti S, Aouchiche H and Oubaziz D 2009 A free-parameter theoretical model for describing the electron elastic scattering in water in the Geant4 toolkit *Radiat. Phys. Chem.* **78** 745–750

[29] Jonah C D, Matheson M S, Miller J R and Hart E J 1976 Yield and decay of the hydrated electron from 100 ps to 3 ns *Journal of Physical Chemistry* **80** 1267-1270

[30] Sumiyoshi T and Katayama M 1982 The yield of hydrated electrons at 30 picoseconds *Chemistry Letters* 1887-1890

[31] Karamitros M and et al. 2014 Diffusion-controlled reactions modeling in Geant4-DNA *Journal of Computational Physics* **274** 841–882

[32] Buxton G V, Greenstock C V, Helman W P and Ross A B 1988 Critical Review of Rate Constants for Reactions of Hydrated Electrons, Hydrogen Atoms and Hydroxyl Radicals (·OH/·O-) in Aqueous Solution *J. Phys. Chem. Ref. Data* **17** 513-586

[33] Kellerer A 1975 *Fandamental of microdosimetry, in The Dosimetry of Ionizing Radiation, vol. 1* Kase K R, Bjarngaard B E and Attix F H Eds. (Academic Press) pp 77-162

[34] Nikjoo H, Goodhead D T, Charlton D E and Paretzke 1989 H G Energy deposition in small cylindrical targets by ultrasoft X-rays *Physics in Medicine and Biology* **34** 691-705,

[35] Nikjoo H and Goodhead D T 1991 Track structure analysis illustrating the prominent role of low energy elec trons in radiobiological eäects of low-LET radiations *Physics in Medicine and Biology* **36** 229-238

[36] ICRU report No. 36 1983 *Microdosimetry* (Maryland: International Commission on Radiation Units and Measurments)

[37] Goodhead D T 1987 *Relationship of microdosimetric techniques to applications in biological systems, in The Dosimetry of Ionising Radiation, vol. 2* Kase K R, Bjarngaard B E and Attix F H Eds. (Orlando, Academic Press) pp 1-89

[38] Charlton D, Nikjoo H and Humm J L 1989 Calculation of initial yields of single- and double-strand breaks in cell nuclei from electrons, protons and alpha particles *Int. J. Radiat. Biol.* **56** 1-19

[39] Nikjoo H, O'Neill P, Terrissol M and Goodhead D T 1999 Quantitative modelling of DNA damage using Monte Carlo track structure method *Radiat. Environ. Biophys.* **38** 31–38

[40] Bernal M A and Liendo J A 2009 An investigation on the capabilities of the PENELOPE MC code in nanodosimetry *Med. Phys.* **36** 620–625

[41] Bernal M A, deAlmeida C E, Sampaio C, Incerti S, Champion C and Nieminen P 2011 The invariance of the total direct DNA strand break yield *Med. Phys.* **38** 4147–4153

[42] Friedland W, Jacob P, Paretzk H G and Stork T 1998 Monte Carlo simulation of the production of short DNA fragments by low-linear energy transfer radiation using higher-order DNA models *Radiat. Res.* **150** 170

[43] Friedland W, Bernhardt P, Jacob P, Paretzke H G and Dingfelder M 2002 Simulation of DNA damage after proton and low LET irradiation *Radiat. Prot. Dosim.* **99** 99–102

[44] Semenenko V A and Stewart R D 2004 A fast Monte Carlo algorithm to simulate the spectrum of DNA damages formed by ionizing radiation *Radiat. Res.* **161** 451–457

[45] Semenenko V A and Stewart R D 2006 Fast Monte Carlo simulation of DNA damage formed by electrons and light ions *Phys. Med. Biol.* **51** 1693–1706

[46] Leslie A G W and Arnott S 1980 Polymorphism of DNA double helices *J. Mol. Biol.* **143** 49-72

[47] Dickerson R E, Drew H R, Conner B N, Wing R M, Fratini A V and Kopka M L 1982 The Anatomy of A-, B-, and Z-DNA *Science* **216** 475- 485

[48] Meylan S, Vimont U, Incerti S, Clairand I and Villagrasa C 2016 Geant4-DNA simulations using complex DNA geometries generated by the DnaFabric tool *Computer Physics Communications* **204** 159–169

[49] Friedland W, Jacob P, Paretzke H G, Merzagora M and Ottolenghi A 1999 Simulation of DNA fragment distributions after irradiation with photons *Radiat. Environ. Biophys.* **38** 39–47

[50] Francis Z, Villagrasa C and Clairand I 2011 Simulation of DNA damage clustering after proton irradiation using an adapted DBSCAN algorithm *Comput. Methods Programs Biomed.* **101** 265–270

[51] Martin R F and Haseltine W A 1981 Range of radiochemical damage to DNA with decay of iodine-125 *Science* **213** 896-898

[52] Terrissol M 1994 Modelling of radiation damage by $^{125}$I on a nucleosome *International Journal of Radiation Biology* **66** 447- 452

[53] Kandaiya S, Lobachevsky P N, D'cunha G and Martin R F 1996 DNA strand breakage by $^{125}$I decay in synthetic oligodeoxynucleotide:1. Fragment distribution and DMSO protection effect *Acta Oncologica* **35** 803-808

[54] Charlton D E and Humm J L 1988 A method of calculating initial DNA strand breakage following the decay of incorporated $^{125}$I *Int. J. Radiat. Biol.* **53** 353-365

[55] Aydogan B, Bolch W E, Swarts S G, Turne J E and Marshal D T 2008 Monte Carlo Simulations of Site-Specific Radical Attack to DNA Bases *Radiation Research* **169** 223–231

[56] Murthy C P, Deeble D J and Sonntag C V 1988 The formation of phosphate end groups in the radiolysis of polynucleotides in aqueous-solution *Z. Naturforsch.* **43** 572-567

[57] Milligan J R, Aguilera J A and Ward J F 1993 Variation of single-strand break yield with scavenger concentration for plasmid DNA irradiated in aqueous solution *Radiation Research* **133** 158-162

[58] Bernal M A, Sikansi D, Cavalcante F, Incerti S, Champion C, Ivanchenko V and Francis Z 2013 An atomistic geometrical model of the B-DNA configuration for DNA–radiation interaction simulations *Computer Physics Communications* **184** 2840–2847

[59] Watanabe R, Rahmanian S and Nikjoo H 2014 Spectrum of base damage induced by electrons and ions-a Monte Carlo track structure simulation calculation *Radiat. Res.* **183** 1–16

[60] Friedland W, Jacob P, Bernhardt P, Paretzke H G and Dingfelder M 2003 Simulation of DNA damage after proton irradiation *Radiation Research* **159** 401-410

[61] Famulari G, Pater P and Enger S A 2017 Microdosimetry calculations for monoenergetic electrons using Geant4-DNA combined with a weighted track sampling algorithm *Phys. Med. Biol.* **62** 5495–5508

[62] Bordage M C, Bordes J, Edel S and et al. 2016 Implementation of new physics models for low energy electrons in liquid water in Geant4-DNA *Physica Medica* **32** 1833-1840

[63] Bernal M A, Bordage M C, Brown J M C and et al. 2015 Track structure modeling in liquid water: A review of the Geant4-DNA very low energy extension of the Geant4 Monte Carlo simulation toolkit *Physica Medica* **31** 861-874

[64] Dingfelder M, Ritchie R H, Turner J E, Friedland W, Paretzke H G and Hamm R N 2008 Comparisons of Calculations with PARTRAC and NOREC: Transport of Electrons in Liquid Water *Radiation Research* **16** 584-594

[65] Cadet J and Wagner J R 2013 DNA Base Damage by Reactive Oxygen Species, Oxidizing Agents, and UV Radiation *Cold Spring Harb. Perspect. Biol.* **5** 12559